\documentclass[sn-basic]{sn-jnl}

\usepackage{graphicx}%
\usepackage{multirow}%
\usepackage{amsmath,amssymb,amsfonts}%
\usepackage{amsthm}%
\usepackage{mathrsfs}%
\usepackage[title]{appendix}%
\usepackage{xcolor}%
\usepackage{textcomp}%
\usepackage{manyfoot}%
\usepackage{booktabs}%
\usepackage{algorithm}%
\usepackage{algorithmicx}%
\usepackage{algpseudocode}%
\usepackage{listings}%
\usepackage{booktabs}%
\usepackage{longtable}
\usepackage{amsmath,amsthm,latexsym,amssymb,enumerate,accents}

\newcommand{\ch}[1]{{\color{black} {#1}}}

\usepackage{tikz,xcolor,hyperref}

\definecolor{lime}{HTML}{A6CE39}
\DeclareRobustCommand{\orcidicon}{%
	\begin{tikzpicture}
	\draw[lime, fill=lime] (0,0) 
	circle [radius=0.16] 
	node[white] {{\fontfamily{qag}\selectfont \tiny ID}};
	\draw[white, fill=white] (-0.0625,0.095) 
	circle [radius=0.007];
	\end{tikzpicture}
	\hspace{-2mm}
}

\foreach \x in {A, ..., Z}{%
	\expandafter\xdef\csname orcid\x\endcsname{\noexpand\href{https://orcid.org/\csname orcidauthor\x\endcsname}{\noexpand\orcidicon}}
}

\begin{document}

\title[stIHC]{Spatial Transcriptomics Iterative Hierarchical Clustering (stIHC): A Novel Method for Identifying Spatial Gene Co-Expression Modules}

\author[1]{\fnm{Catherine} \sur{Higgins}\href{https://orcid.org/0009-0003-8718-9335}{\orcidicon}}\email{catherine.higgins1@ucdconnect.ie}

\author*[2]{\fnm{Jingyi Jessica} \sur{Li}\href{https://orcid.org/0000-0002-9288-5648}{\orcidicon}}\email{ jli@stat.ucla.edu}

\author[1]{\fnm{Michelle} \sur{Carey}\href{https://orcid.org/0000-0002-5603-4264}{\orcidicon}}\email{michelle.carey@ucd.ie}

\affil[1]{\orgdiv{School of Mathematics and Statistics}, \orgname{University College Dublin}, \orgaddress{ \country{Ireland}}}

\affil[2]{\orgdiv{Department of Statistics and Data Science}, \orgname{University of California, Los Angeles}, \orgaddress{\country{United States of America}}}

\abstract{Recent advancements in spatial transcriptomics technologies allow researchers to simultaneously measure RNA expression levels for hundreds to thousands of genes while preserving spatial information within tissues, providing critical insights into spatial gene expression patterns, tissue organization, and gene functionality. However, existing methods for clustering spatially variable genes (SVGs) into co-expression modules often fail to detect rare or unique spatial expression patterns. To address this, we present spatial transcriptomics iterative hierarchical clustering (stIHC), a novel method for clustering SVGs into co-expression modules, representing groups of genes with shared spatial expression patterns. Through three simulations and applications to spatial transcriptomics datasets from technologies such as 10x Visium, 10x Xenium, and Spatial Transcriptomics, stIHC outperforms clustering approaches used by popular SVG detection methods, including SPARK, SPARK-X, MERINGUE, and SpatialDE. Gene Ontology enrichment analysis confirms that genes within each module share consistent biological functions, supporting the functional relevance of spatial co-expression. Robust across technologies with varying gene numbers and spatial resolution, stIHC provides a powerful tool for decoding the spatial organization of gene expression and the functional structure of complex tissues.}


\maketitle
\newpage
\section{Introduction}\label{sec1}
Recent advancements in spatial transcriptomics (ST) technologies have enabled the measurement of gene expression levels while preserving spatial information within tissues. These technologies include sequencing-based platforms such as Spatial Transcriptomics \citep{shah2016situ}, 10x Visium \citep{10xGenomics}, and Slide-seq \citep{rodriques2019slide,stickels2021highly}, as well as imaging-based platforms such as 10x Xenium \citep{janesick2023high}, MERFISH \citep{chen2015spatially}, and seqFISH \citep{lubeck2014single}, which allow the quantification of hundreds to thousands of genes in relation to their spatial location.

Many methods have been developed to identify spatially variable genes (SVGs), which exhibit significant spatial expression variability across a tissue. For a comprehensive overview of these methods, see \cite{yan2024categorization}. Despite substantial progress in identifying SVGs, there remains a lack of methods for clustering these genes into co-expression modules—groups of genes with similar spatial expression patterns. Since biological pathways and processes are often driven by the coordinated activity of multiple genes \citep{van2018gene,zinani2022regulatory,russell2023gene}, identifying co-expression modules can enhance our understanding of the spatial organization and function of tissues. For example, this approach can help identify regions in tumors with unique gene expression patterns and functional roles \citep{yuan2024heartsvg}. 

\begin{figure}[h!]
\begin{center}
 {{\includegraphics[width=1.05\linewidth]{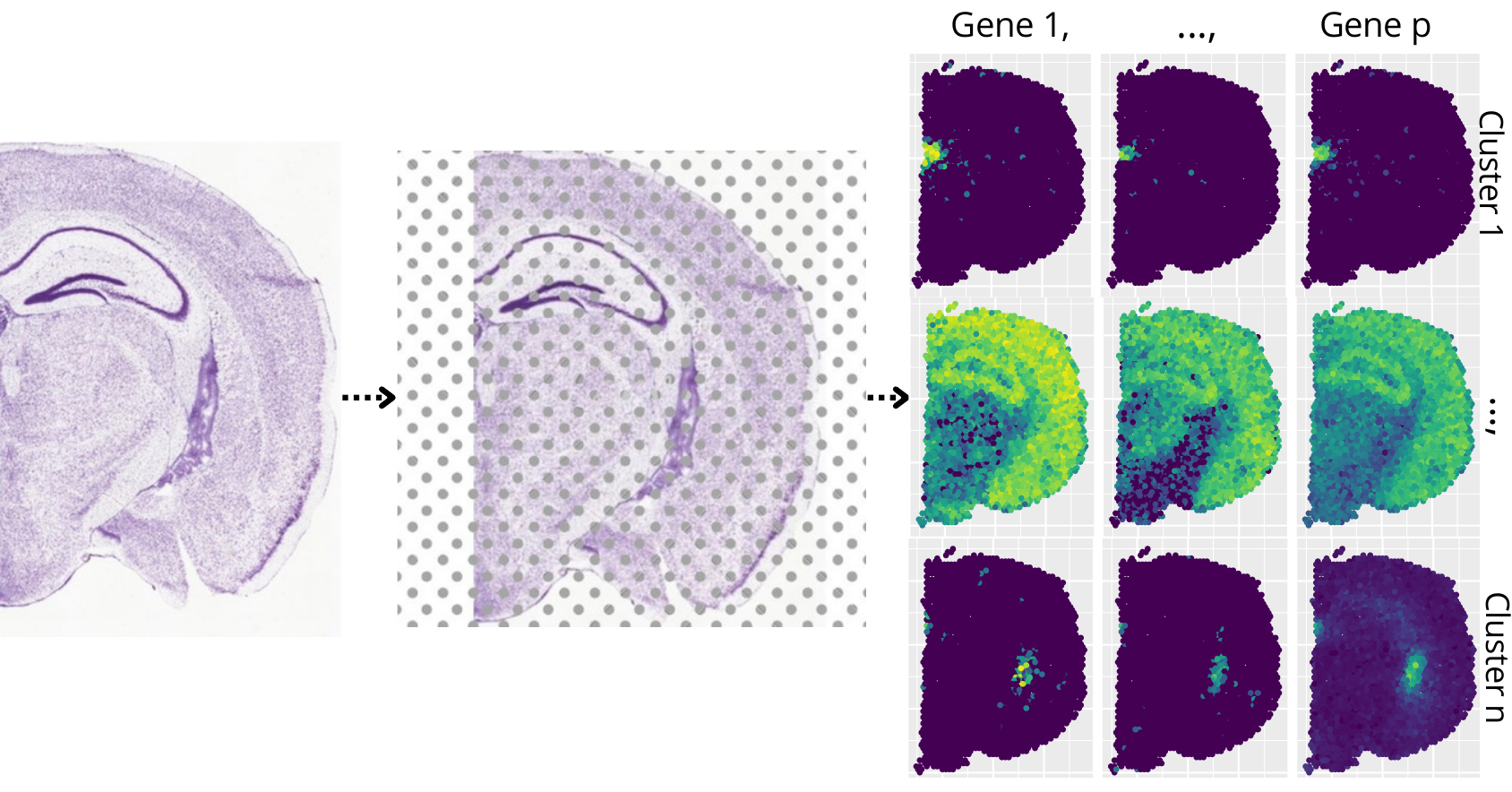} }}%
 \caption{Workflow of spatial transcriptomics analysis. (Left) Tissue section, adapted from the Allen Mouse Brain Atlas (mouse.brain-map.org) \citep{allen1}. (Middle) Spatial transcriptomics grid overlay, representing spatially resolved gene expression measurements. (Right) Identified gene co-expression modules, illustrating groups of genes with distinct spatial expression patterns. \label{Figintro}}   
\end{center}

 \end{figure}

Numerous studies have focused on clustering cells or spatial spots in ST data; for an overview of these methods, see \cite{cheng2023benchmarking}. However, these methods differ fundamentally from our proposed method. Spot-based methods cluster spatial locations, typically corresponding to cell types, states, or functional regions within the tissue. In contrast, our focus is on gene co-expression modules, clustering at the gene level. Here, clusters represent groups of genes exhibiting similar spatial expression patterns across all spatial spots. Figure \ref{Figintro} illustrates this distinction. The left panel depicts the tissue slice, the middle panel overlays a grid representing the spatial spots where gene abundance is measured using ST technologies, and the right panel shows the resulting gene expression measurements for nine genes across the tissue slice. As shown, many genes exhibit similar spatial expression patterns; in this example, the nine genes display three distinct patterns. The aim of our method is to cluster genes with similar spatial behavior across the entire tissue into meaningful groups. This shift from spot-based analysis to gene-based analysis emphasizes uncovering relationships among genes rather than identifying cell types or spatial domains.

Some existing methods for detecting SVGs in ST data incorporate gene clustering as a secondary step, but these methods primarily aim to identify SVGs rather than explicitly detect spatial gene co-expression modules. Typically, these methods identify SVGs using a statistic calculated for each gene and subsequently group the detected SVGs into clusters. For example, SpatialDE \citep{svensson2018spatialde} and SPARK \citep{sun2020statistical} use Gaussian process regression to decompose the expression levels of each gene across tissue into a spatial component and an independent noise term, identifying a gene as an SVG if the spatial component is statistically significant. The effectiveness of Gaussian process regression depends heavily on how accurately the selected spatial covariance matrix (derived from a chosen kernel) models spatial patterns \citep{yan2024categorization}. After detecting SVGs, SpatialDE applies an extended Gaussian mixture model \citep{mclachlan1988mixture} to cluster SVGs, utilizing the same Gaussian-process-based prior employed during SVG detection. Another method SPARK-X \citep{zhu2021spark} identifies SVGs by testing the independence between two spot similarity matrices: one based on the expression levels of each gene and the other based on kernel-transformed spatial locations. A gene is identified as an SVG if the null hypothesis of independence is rejected. Like SpatialDE and SPARK, SPARK-X’s effectiveness relies on the ability of the chosen kernel to capture spatial patterns. Following SVG detection, both SPARK and SPARK-X apply a log transformation to the raw count data, scaling the counts of each SVG to zero mean and unit variance across spots, and then perform hierarchical agglomerative clustering to group SVGs into co-expression modules. MERINGUE \citep{miller2021characterizing}, a graph-based method, constructs a neighborhood graph of spatial spots using Delaunay triangulation, assigning binary edge weights (\(w_{ij} = 1\) if two spots are connected, \(w_{ij} = 0\) otherwise). It detects SVGs based on Moran’s I, transforming it into a z-statistic and calculating a one-sided p-value for each gene. To cluster SVGs, MERINGUE calculates a spatial cross-correlation index for each SVG pair, generating a spatial cross-correlation matrix that is then used in hierarchical clustering to group SVGs into gene co-expression modules. While these methods incorporate clustering, their primary focus remains on identifying SVGs, and their clustering steps often rely on the same assumptions or data transformations used during SVG detection.

Certain spatial patterns may involve a large number of co-expressed genes, while others may involve only a few or even a single gene. This pattern heterogeneity can arise from factors such as pathological differences within tissue regions (e.g., tumor versus normal areas in cancer) or the presence of diverse or rare cell types \citep{sun2020statistical,acharyya2022spacex}. Consequently, gene co-expression modules may be imbalanced clusters, with some containing many genes while others are small or consist of a single gene. Standard clustering methods, such as mixture-model-based clustering \citep{mclachlan1988mixture} and hierarchical clustering \citep{eisen1998cluster}, often lack the flexibility to capture such imbalanced clusters effectively. Although this challenge has been addressed in other contexts, such as time-course gene expression \citep{carey2016correlation, higgins2024funIHC}, it has not, to our knowledge, been specifically addressed in the context of ST data.

We present spatial transcriptomics iterative hierarchical clustering (stIHC), a novel two-step method designed to identify gene co-expression modules, including those with imbalanced sizes. In the first step, stIHC models the expression levels of SVGs across the spatial domain using a generalized penalized regression framework \citep{sangalli2013spatial, wilhelm2016generalized}, enabling efficient modeling of spatial variation over complex domains, including those with intricate geometries. This step is essential because gene expression data typically exhibit substantial noise while also following an intrinsically smooth spatial pattern. By modeling a gene's expression levels at spatial spots as a smooth, continuous function rather than discrete values, it becomes possible to evaluate gene similarity across the entire tissue while accounting for spatial dependencies. This approach integrates relationships between adjacent spatial spots rather than treating a gene's expression levels at different spots as independent observations. In the second step, stIHC clusters SVGs into co-expression modules based on the estimated model coefficients. Recognizing that standard clustering methods often struggle with imbalanced gene clusters and fail to capture rare spatial expression patterns, stIHC employs the recently developed functional iterative hierarchical clustering (funIHC) \citep{higgins2024funIHC}. This approach is specifically designed to handle imbalanced clusters, ensuring robust performance in identifying gene modules of any size that exhibit similar spatial expression patterns across a tissue slice.

We evaluate the effectiveness of stIHC across three simulated and four ST datasets generated using 10x Visium, 10x Xenium, and Spatial Transcriptomics technologies. Its performance is compared against the clustering approaches used by SpatialDE \citep{svensson2018spatialde}, SPARK \citep{sun2020statistical}, SPARK-X \citep{zhu2021spark}, and MERINGUE \citep{miller2021characterizing} for grouping SVGs into co-expression modules. The performance of stIHC is assessed using the adjusted Rand index (ARI) and the Davies-Bouldin index (DBI) to determine its ability to identify coherent spatial gene expression modules and capture rare or unique spatial patterns, which are often missed by the clustering approaches employed in the four SVG detection methods. Additionally, we examine whether the identified co-expression modules are not only spatially coherent but also functionally relevant by analyzing their biological annotations and evaluating alignment with the known roles of the corresponding anatomical regions.

The contributions of this paper are as follows: First, we introduce stIHC, a novel clustering method for ST data that identifies spatial co-expression gene modules, including those with unique spatial expression patterns often overlooked or merged into larger clusters by existing methods. Second, our method is data-driven and parameter-lean, requiring no user-defined input parameters. Finally, by revealing co-expression modules, our method provides valuable insights into the spatial organization of gene expression, establishing a powerful framework for exploring tissue structure and spatially coordinated biological processes.

The remainder of the paper is organized as follows: Section \ref{sec2} introduces our proposed clustering method, stIHC. Section \ref{sec3} presents simulation studies evaluating the performance of stIHC compared to existing approaches. Section \ref{sec4} explores the functional annotation of gene clusters identified by stIHC. Section \ref{sec5} demonstrates the application of stIHC to ST data from the mouse olfactory bulb. Finally, Section \ref{sec6} concludes with a summary and discussion.

\section{The stIHC Methodology}\label{sec2}

ST data comprises gene expression levels measured at \(n\) distinct spatial locations, or ``spots," within a tissue sample. For each spot \(j = 1, \dots, n\), its 2D spatial location is denoted as \(s_j = (s_{j1}, s_{j2})^\top \in \mathbb{R}^2\). The expression level of gene \(i = 1, \dots, G\) at spot \(j\) is represented by \(y_{ji} \in \mathbb{R}\). Our method, stIHC, identifies gene co-expression modules through a two-step process: (1) Modeling gene expression across the 2D tissue slice using a generalized penalized regression framework \citep{sangalli2013spatial,wilhelm2016generalized}, and (2) Clustering the basis coefficients obtained from the spatial model using funIHC, a functional iterative hierarchical clustering algorithm \citep{higgins2024funIHC}.

\subsection{Step 1: Modeling Gene Expression in a 2D Tissue Slice}

Consider \(n\) locations \(\mathbf{s} = (s_1, \dots, s_n)\) within a 2D tissue slice \(\Omega \subseteq \mathbb{R}^2\). At location \(s_j\), the observed expression level of gene $i$ is denoted by \(y_{ji} \in \mathbb{R}\). We assume the response variable \(y_{ji}\) follows a distribution within the exponential family with mean \(\mu_i\). The exponential family includes many common distributions, such as Poisson, Binomial, Gamma, and Normal \citep{mccullagh1989}. Thus, this framework can be used for modeling raw gene expression counts (e.g., Poisson) or normalized gene expression (e.g., Normal).

To begin, the slice of interest \(\Omega\) is partitioned into smaller regions using Delaunay triangulation \citep{shewchuk1996triangle,shewchuk2002delaunay}, an efficient algorithm for handling non-regular geometries, such as the irregular tissue shape shown in the left panel of Figure~\ref{Figintro}. Locally supported polynomial functions are defined over these triangles, providing a set of basis functions. Let \(\boldsymbol{\phi}(\mathbf{s}) = (\phi_1(\mathbf{s}), \dots, \phi_K(\mathbf{s}))^\top\) represent an \(n \times K\) matrix containing \(K\) piecewise linear basis functions evaluated at the locations \(\mathbf{s}\). Using these basis functions, we model \(\mu_i\) via a generalized penalized regression framework \citep{sangalli2013spatial, wilhelm2016generalized}. 

Let \(g\) be a continuously differentiable and strictly monotone canonical link function, such as \(g(\mu) = \log(\mu)\) (Poisson) or \(g(\mu) = \mu\) (Normal), and let \(f\) represent a smooth spatial field over \(\Omega\). Then:
\begin{equation}
    g(\mu_i) = f_i(\mathbf{s}) = \boldsymbol{\phi}(\mathbf{s}) \boldsymbol{c_i},
\end{equation}
where \(\boldsymbol{c_i}\) are the coefficients of the basis function expansion approximating the behavior of \(g(\mu_i)\). 

The coefficients \(\boldsymbol{c_i}\) are estimated by maximizing a penalized log-likelihood functional:
\begin{equation}\label{likeli}
    \mathcal{L}_s(\boldsymbol{c_i}) = \sum_{j=1}^n l(y_{ji}; \boldsymbol{c_i}) - \lambda \int_\Omega (\Delta f_i(\mathbf{s}))^2 \, d\mathbf{s},
\end{equation}
where \(l(\cdot)\) is the log-likelihood, \(\lambda > 0\) is a smoothing parameter, and the Laplacian \(\Delta f_i = \frac{\partial^2 f_i}{\partial s_1^2} + \frac{\partial^2 f_i}{\partial s_2^2}\) measures the local curvature of the spatial field \(f_i\). Increased values of the smoothing parameter $\lambda$ yield smoother estimates of $f$, whereas decreased values of $\lambda$ give estimates that more closely fit the data. The Laplacian serves as the optimal penalization choice for mitigating the effects of noise present in the data \citep{sangalli2013spatial}. Nevertheless, if the user has domain-specific insights that require penalizing deviations from a more complex partial differential equation, such adjustments are possible; see \cite{sangalli2021spatial} for further details.  
Using a penalized iterative reweighted least squares algorithm \citep{wilhelm2016generalized}, we solve the optimization problem in (\ref{likeli}) to estimate the coefficients \(\boldsymbol{c_i}\) for each gene $i$ for fixed $\lambda$. The optimal smoothing parameter \( \lambda \) is estimated by minimizing the generalized cross-validation criterion, as proposed by \cite{craven1978smoothing}, across all genes, thereby determining a unified \( \lambda \) that ensures consistent penalization throughout the entire gene set. Let \( \mathbf{C} \) represent the \( G \times K \) matrix of coefficients corresponding to the optimal \( \lambda \) configuration, where $\mathbf{C} = (\boldsymbol{c}_1, \dots, \boldsymbol{c}_G)^\top$
denotes the complete set of coefficients for all \( G \) genes.

\subsection{Step 2: Clustering of Genes Based on Spatial Expression Patterns}

Using the resulting basis coefficients for all genes, the distance metric \(d_{i,j}\), quantifying the dissimilarity between two genes $i$ and $j$, is computed as one minus the Spearman correlation \(\rho_{i,j}\) between $\boldsymbol{c}_i$ and $\boldsymbol{c}_j$, i.e., the \(i^{\text{th}}\) and \(j^{\text{th}}\) rows of \(\mathbf{C}\). The clustering procedure proceeds as follows:

\begin{enumerate}
\item Define \(\alpha_{\min}\) and \(\alpha_{\max}\) as the minimum and maximum values of the Spearman correlation computed across all possible gene pairs \(\{\rho_{i,j}\}_{i,j=1}^{G}\). Construct a grid spanning \([\alpha_{\min}, \alpha_{\max}]\), comprising \(U\) equally spaced values. For each \(\alpha_u\), where \(u = 1, \dots, U\):
\begin{enumerate}[i]
\item \textbf{Cluster:} Perform hierarchical clustering on the genes using the distance metric \(d_{i,j}\) with a threshold of \(1 - \alpha_u\). Use the average linkage method to determine the dissimilarity between clusters. Denote the resulting number of clusters as \(P\).
\item \textbf{Merge:} Let \(\mu_p\) represent the center of cluster \(p\) (\(p = 1, \dots, P\)). Treat each \(\mu_p\) as a new gene and apply the same criterion as in Step i to decide whether any cluster centers \(\mu_p\) should be merged. If merging is identified, consolidate the respective clusters into a single one, resulting in \(R\) clusters, where \(R \leq P\).
\item \textbf{Prune:} For each cluster \(r = 1, \dots, R\), if the Spearman correlation \(\rho_{i,r}\) between \(\mu_r\) and the \(i^{\text{th}}\) gene within falls below \(\alpha_u\), remove gene \(i\) from cluster \(r\). All removed genes are allocated into single-gene clusters. This results in a total of \(R + S\) clusters, where \(S\) is the number of removed genes.
\item \textbf{Repeat Steps ii and iii:} Continue merging and pruning until the cluster assignment converges (see Appendix \ref{con} for details).
\item \textbf{Repeat Step ii:} Repeat the merging step until all between-cluster-center Spearman correlations are less than \(\alpha_u\).
\end{enumerate}
\item Determine the optimal value \(\alpha_{\text{opt}}\) from \(\{\alpha_1, \dots, \alpha_U\}\) by maximizing the average silhouette value across genes. The silhouette value for the \(i^{\text{th}}\) gene is defined as:
\[
\text{sil}_i = \frac{b_i - a_i}{\max(a_i, b_i)},
\]
where \(a_i\) is the average distance from the \(i^{\text{th}}\) gene to other genes in the same cluster, and \(b_i\) is the minimum average distance from the \(i^{\text{th}}\) gene to genes in any other cluster. The silhouette value ranges from \(-1\) to \(1\), with higher values indicating stronger similarity within clusters and greater dissimilarity between clusters.
\end{enumerate}

\section{Simulations}\label{sec3}
We conducted three simulations to evaluate the clustering performance of stIHC for ST data in comparison to the clustering approaches used in four SVG detection methods: SpatialDE \citep{svensson2018spatialde}, SPARK \citep{sun2020statistical}, SPARK-X \citep{zhu2021spark}, and MERINGUE \citep{miller2021characterizing}. Implementation details are provided in Appendix \ref{secA2}. These simulations were based on the 10x Visium sagittal mouse brain slice dataset, available in the \texttt{Seurat} R package \citep{satija2015spatial}. The first simulation evaluates performance under an ideal scenario with balanced, equally sized clusters (Section \ref{balsim}). The second simulation tests performance with imbalanced cluster sizes (Section \ref{simimbal}). The third simulation examines performance on sparse, imbalanced clusters to mimic the real ST data analyzed in Section \ref{sec5} (Section \ref{secsmallsim}). To assess the stability of stIHC, we repeated the simulations 100 times and observed that the method consistently produced identical clustering results across all runs. This demonstrates that stIHC is a robust method. We also note that stIHC does not introduce variability due to random initialization or stochastic processes, ensuring reproducible clustering on a particular dataset. Moreover, Appendix \ref{time} presents the execution time in seconds for each method and simulation scenario. stIHC consistently demonstrates efficient computational performance across the three simulation settings.

\subsection{Clustering Performance with Equally Sized Spatial Co-expression Modules}\label{balsim}

\begin{figure}[h!]
    \centering
    \includegraphics[width=1\linewidth]{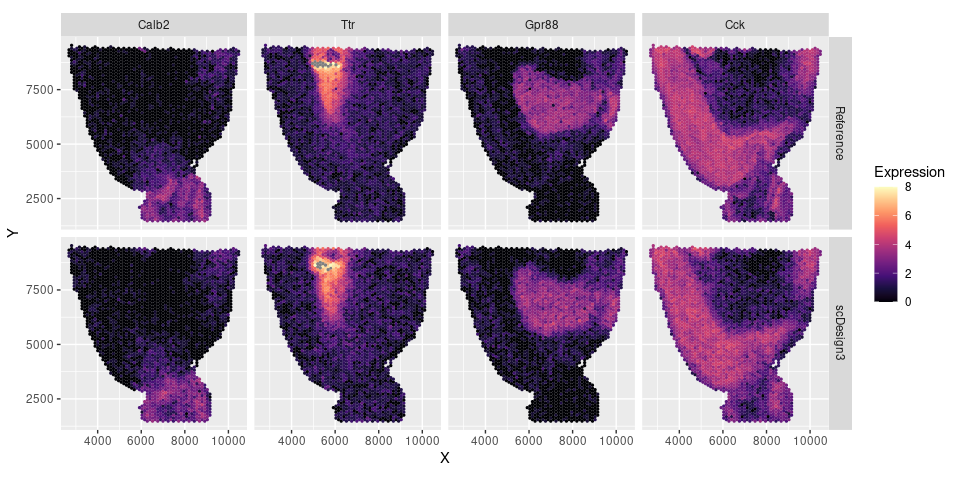}
    \caption{Four genes from the 10x Visium sagittal mouse brain slice dataset representing the four clusters (top row). Example of one simulated gene from each cluster generated using \texttt{scDesign3} (bottom row). \label{Fig1}}
\end{figure}

This first simulation evaluates clustering performance when all gene modules have clearly distinguishable spatial expression patterns and an equal number of genes in each module. To represent distinct spatial patterns, we selected four unique gene expression patterns from the 10x Visium sagittal mouse brain slice dataset, each serving as the representative pattern for a different module. For a visual representation of these spatial expression patterns, refer to the top row of Figure~\ref{Fig1}. 

We simulated the spatial expression patterns of 100 genes from the four representative patterns (25 genes generated per pattern) using \texttt{scDesign3} \citep{song2024scdesign3}. For illustration, an example of one simulated gene from each cluster is displayed in the bottom row of Figure~\ref{Fig1}. Full details of the data simulation process are provided in Appendix~\ref{secA1}.

We applied five clustering methods: our proposed method, stIHC, along with the clustering approaches in SpatialDE, SPARK, SPARK-X, and MERINGUE. We evaluated the performance of each method using the Adjusted Rand Index (ARI) \citep{hubert1985comparing}, which ranges from -1 to 1, with 1 indicating complete agreement between the clusters and the ground-truth gene modules. All methods successfully partitioned the 100 genes into their respective clusters, achieving a perfect clustering solution with an ARI of 1. These results indicate that the methods effectively clustered genes with similar spatial expression patterns in this balanced scenario with distinct clusters.

\subsection{Clustering Performance with Imbalanced Spatial Co-expression Modules}\label{simimbal}

The first simulation provided a baseline assessment of performance under ideal conditions with balanced gene modules. However, real-world ST data may have gene modules of varying sizes. To simulate this more realistic scenario, we adjusted the dataset from the first simulation to create imbalanced gene modules. Specifically, we simulated 6 genes in the first module with expression resembling \textit{Calb2}, 2 genes in the second module resembling \textit{Ttr}, 16 genes in the third module resembling \textit{Gpr88}, and 25 genes in the fourth module resembling \textit{Cck} (see Figure~\ref{Fig1}). Performance was assessed using the Adjusted Rand Index (ARI) \citep{hubert1985comparing} and the Davies-Bouldin Index (DBI) \citep{davies1979cluster}, which ranges from \([0, \infty)\), with lower values indicating better clustering quality. Results are summarized in Table~\ref{tabimbal}.

When applied to this imbalanced dataset, stIHC was the only method to correctly identify all four modules along with the correct number of genes in each module. It achieved an ARI of 1 and the lowest DB index of 0.49. In contrast, SPARK, SPARK-X, and MERINGUE identified the same three clusters (sizes: 6, 18, and 25 genes) and failed to detect the smallest cluster of just two genes, resulting in an ARI of 0.94 and a DB index of 0.54. SpatialDE performed the worst, identifying four clusters but completely misrepresenting the spatial patterns, resulting in cluster sizes of 16, 13, 8, and 12 genes, with a poor ARI of 0.11 and a high DB index of 4.54. These results highlight the limitations of SPARK, SPARK-X, MERINGUE, and SpatialDE, which tend to merge smaller clusters into larger ones. In contrast, stIHC effectively captures and preserves rare spatial patterns.

\begin{table}[ht]
\caption{Adjusted Rand Index (ARI) and Davies-Bouldin Index (DBI) for each method on the imbalanced simulation.}\label{tabimbal}
\normalsize
\centering
\begin{tabular*}{\textwidth}{@{\extracolsep{\fill}}lccccc}
\toprule
Metric & \textbf{stIHC} & SPARK & SPARK-X & MERINGUE & SpatialDE \\
\midrule
ARI & \textbf{1.00} & 0.94 & 0.94 & 0.94 & 0.11  \\
DBI  & \textbf{0.49} & 0.54 & 0.54 & 0.54 & 4.54  \\
\bottomrule
\end{tabular*}
\end{table}

\begin{figure}[h!]
\centering
\includegraphics[width=0.7\linewidth]{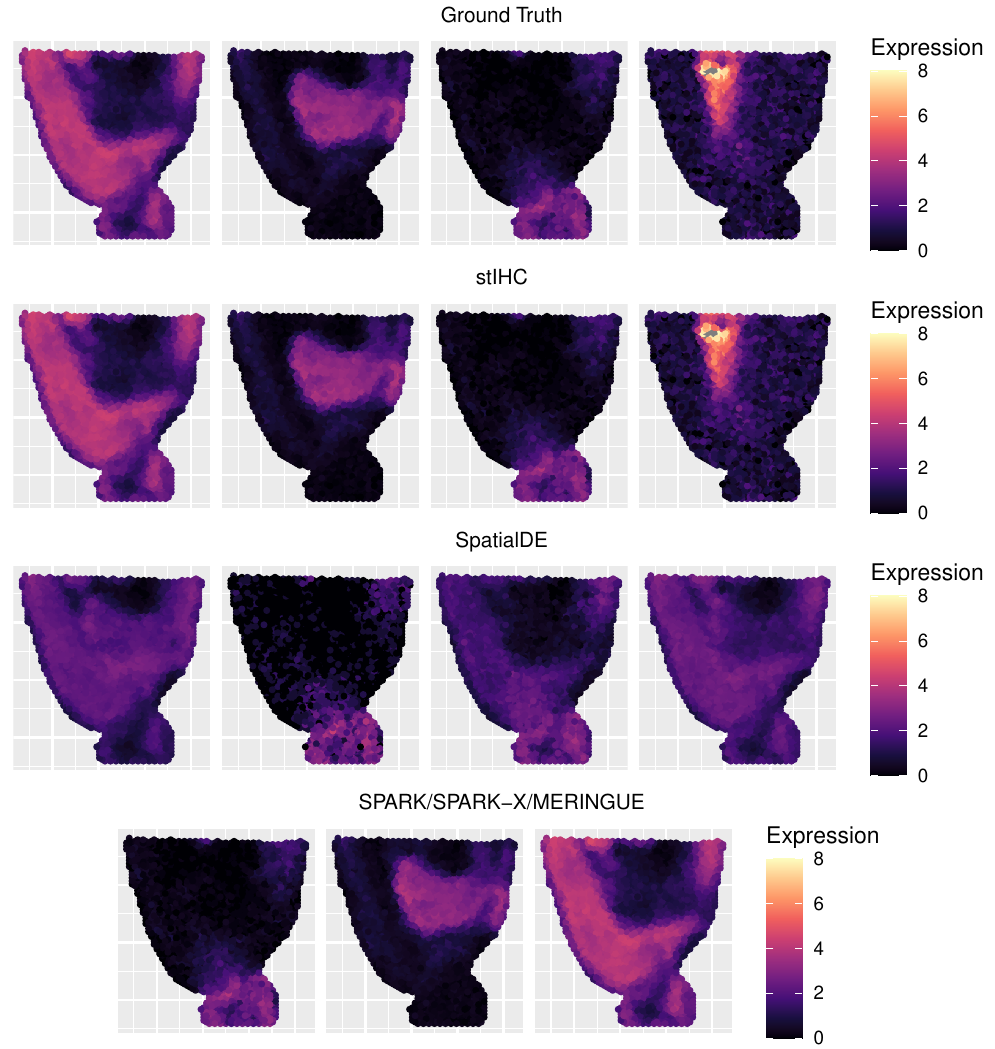}
\caption{The ground-truth mean spatial expression patterns for each gene module in the imbalanced simulation (top row), followed by the mean spatial expression patterns identified by stIHC (second row), SpatialDE (third row), and SPARK, SPARK-X, and MERINGUE (bottom row). \label{Fig3}}
\end{figure}

We further evaluated the clustering methods by comparing the mean spatial expression patterns of the clusters identified by each method to the ground-truth patterns used in the simulation. As shown in Figure~\ref{Fig3}, methods that failed to capture the ground-truth gene modules produced distorted spatial expression patterns. The ground-truth spatial expression patterns of gene modules are presented in the top row, followed by the mean patterns identified by stIHC (second row), SpatialDE (third row), and SPARK, SPARK-X, and MERINGUE (bottom row).

Specifically, stIHC accurately identified the correct clusters, maintaining clear, consistent, and distinct mean spatial expression patterns for each cluster. It was the only method to correctly capture the spatial pattern of \textit{Ttr}. SpatialDE partially captured the patterns of \textit{Calb2} and \textit{Cck} but failed to capture the patterns of \textit{Ttr} and \textit{Gpr88}. SPARK, SPARK-X, and MERINGUE failed to capture the spatial pattern of \textit{Ttr} but adequately captured the patterns of \textit{Calb2}, \textit{Gpr88}, and \textit{Cck}. These findings demonstrate that SpatialDE, SPARK, SPARK-X, and MERINGUE are limited in their ability to capture the spatial expression patterns of rare or small gene modules, whereas stIHC effectively identifies these patterns.

\newpage
\subsection{Clustering Performance with Sparse Spatial Resolution and Imbalanced Spatial Co-expression Modules}\label{secsmallsim}%
This simulation evaluates clustering performance on a sparse version of the imbalanced dataset described in Section \ref{simimbal}, where 260 spatial locations were randomly sampled. This sparse spatial resolution mimics the real dataset analyzed in Section \ref{sec5} and is designed to test the stIHC's effectiveness under reduced spatial resolution. We maintained the same imbalanced module sizes as in Section \ref{simimbal}: 6 genes in the first module reflecting the spatial expression pattern of \textit{Calb2}, 2 genes in the second module reflecting the pattern of \textit{Ttr}, 16 genes in the third module representing \textit{Gpr88}, and 25 genes in the fourth module representing \textit{Cck}. Performance was assessed using the Adjusted Rand Index (ARI) and the Davies-Bouldin Index (DBI), with results summarized in Table \ref{tabsmall}.

\begin{figure}[h!]
 \centering
 \includegraphics[scale=.53]{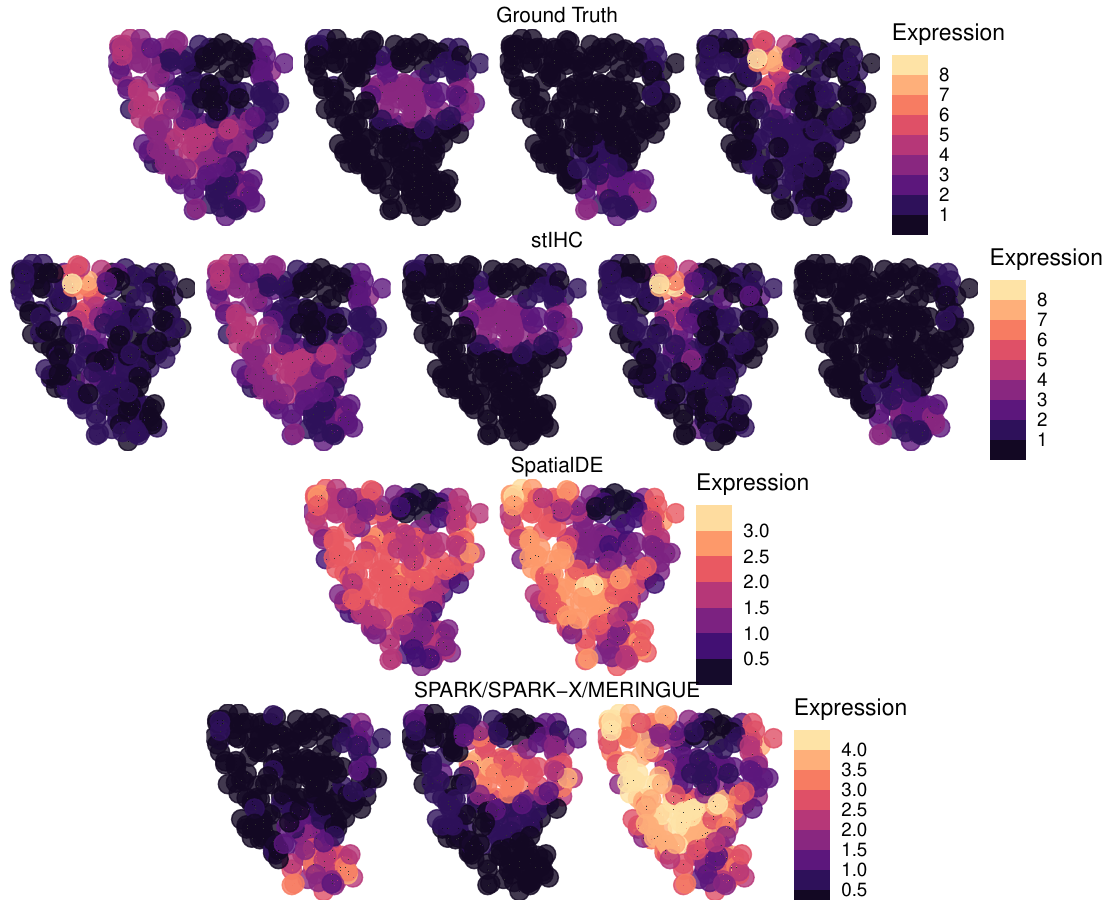}
 \caption{The ground truth mean spatial expression patterns for each cluster in the sparse imbalanced simulation (top row), followed by the mean spatial expression patterns identified by stIHC (second row), SpatialDE (third row), and SPARK, SPARK-X, and MERINGUE (bottom row).\label{smallsimmeans}}
\end{figure}

When applied to this sparse imbalanced dataset, stIHC identified five clusters with sizes 6, 1, 1, 16, and 25. It correctly recovered the number of genes in the \textit{Calb2}, \textit{Cck}, and \textit{Gpr88} module but split the smallest module (\textit{Ttr}) into two separate clusters, each containing a single gene. Despite this, stIHC achieved the highest ARI of 0.99 and the lowest DBI of 0.37 among all methods examined. Notably, stIHC was the only method capable of preserving the spatial expression pattern of the smallest module, as evidenced by comparing its mean spatial expression patterns to the ground truth in Figure \ref{smallsimmeans}.

In contrast, SPARK, SPARK-X, and MERINGUE identified only three clusters with sizes of 6, 18, and 25 genes, consistent with the results from Section \ref{simimbal}. These methods failed to detect the smallest module of two genes, leading to a lower ARI of 0.94 and a higher DBI of 0.54. While the mean spatial expression patterns for the identified clusters retained the patterns of \textit{Calb2}, \textit{Cck}, and \textit{Gpr88}, the spatial pattern of the smallest module (\textit{Ttr}) was entirely missed, as shown in Figure \ref{smallsimmeans}. 

SpatialDE performed the worst, identifying only two clusters with sizes 31 and 18. This resulted in a significantly lower ARI of 0.05 and a much higher DBI of 4.45. The mean spatial expression patterns produced by SpatialDE were significantly distorted compared to the ground truth, highlighting its limitations in handling sparse spatial resolution and imbalanced spatial co-expression modules (Figure \ref{smallsimmeans}).

\begin{table}[ht]
\caption{Adjusted Rand Index (ARI) and Davies-Bouldin Index (DBI) for each method on the sparse imbalanced simulation with 260 sampled locations.}\label{tabsmall}
\normalsize
\centering
\begin{tabular*}{\textwidth}{@{\extracolsep{\fill}}lccccc}
\toprule
Metric & \textbf{stIHC} & SPARK & SPARK-X & MERINGUE & SpatialDE \\
\midrule
ARI & \textbf{0.99} & 0.94 & 0.94 & 0.94 & 0.05  \\
DBI & \textbf{0.37} & 0.54 & 0.54 & 0.54 & 4.45  \\
\bottomrule
\end{tabular*}
\end{table}

\section{stIHC Identifies Functional Gene Modules by Clustering Spatial Expression Patterns}\label{sec4}

To further validate stIHC, we applied it to three real ST datasets: 10x Visium mouse brain, 10x Xenium mouse brain, \ch{and 10x Visium human lung cancer} (Sections \ref{10xV}, \ref{10xX} \ch{and \ref{10xvisBC}, respectively}). Our goal was to determine whether genes with similar spatial expression patterns also share biological functions. Specifically, we evaluated the biological coherence and functional distinctiveness of each detected co-expression module. For the 10x Visium mouse brain and 10x Xenium mouse brain datasets, we assessed the spatial expression pattern similarity within clusters, alignment with known tissue structure, and gene ontology (GO) enrichment analysis. We also examined the consistency of results across the two platforms and tissue orientations. For the 10 Visium human lung cancer dataset, we evaluated the similarity of spatial expression patterns within clusters and performed GO enrichment analysis.

Each dataset consisted of 50 curated genes identified as SVGs. This selection ensures high confidence in the SVGs and removes potential biases from SVG detection methods, allowing for a focused evaluation of clustering performance. While our method supports datasets of any size, we limited this analysis to curated SVGs for consistency. Details of SVG selection are provided in Appendix \ref{secA1}.

\subsection{10x Visium Mouse Brain Dataset}\label{10xV}

\begin{figure}[h!]
\begin{center}
{{\includegraphics[width=1.05\linewidth]{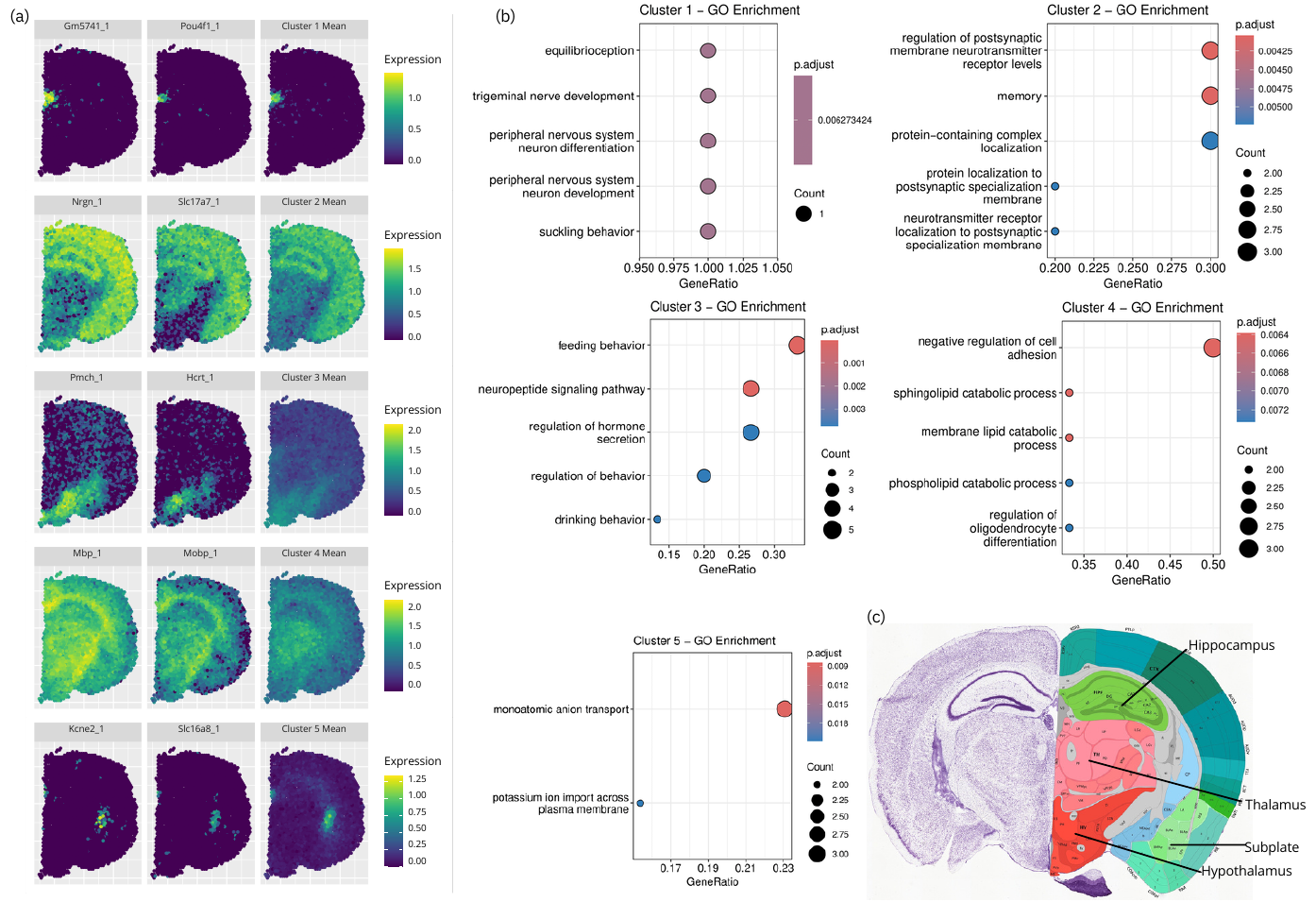}}}%
\caption{(a) Mean spatial expression patterns of the five clusters with two representative genes per cluster in the 10x Visium mouse brain dataset. (b) GO enrichment terms for the clusters. (c) Anatomical regions from the Allen Mouse Brain Atlas and Allen Reference Atlas (mouse.brain-map.org and atlas.brain-map.org) \citep{allen1,allen2}. \label{Fig5}}
\end{center}
\end{figure}

The 10x Visium mouse brain dataset was preprocessed to include 50 curated SVGs (Appendix \ref{secA1}). Using stIHC, we identified five co-expression modules containing 2, 10, 16, 7, and 15 genes, respectively. Figure \ref{Fig5}(a) shows the mean spatial expression pattern for each cluster, along with two representative genes per cluster. These spatially distinct patterns demonstrate that stIHC effectively captures the inherent spatial structure of the data.

To evaluate the biological relevance of these clusters, we performed GO enrichment analysis using the \texttt{clusterProfiler} R package \citep{yu2012clusterprofiler}. Each cluster was enriched for unique biological processes (Figure \ref{Fig5}(b)). Metrics such as p.adjust (adjusted p-value), gene ratio (the proportion of genes associated with a GO term), and count (the number of associated genes) confirm the distinct functional identities of the clusters. Full gene lists are provided in Appendix \ref{fullgenelists}.

We compared the spatial expression patterns of these clusters with anatomical regions from the Allen Mouse Brain Atlas (Figure \ref{Fig5}(c)). Each cluster corresponded to a specific brain region. 
\begin{itemize}
    \item Cluster 1 (2 genes): Thalamus, aligning with pathways related to sensory and motor signal relay \citep{shine2023impact}.
    \item Cluster 2 (10 genes): Hippocampus, including the dentate gyrus and Ammon’s horn, with GO terms highlighting memory processes \citep{pang2019ammon,borzello2023assessments}.
    \item Cluster 3 (16 genes): Hypothalamus, associated with hormone regulation and production \citep{chrousos1995hypothalamic}.
    \item Cluster 4 (7 genes): Thalamus and hippocampus, enriched for catabolic processes.
    \item Cluster 5 (15 genes): Cortical subplate, linked to migration and ion transport \citep{luhmann2009subplate}.
\end{itemize}
These findings confirm that the spatially distinct gene modules identified by stIHC align with the functional organization of the tissue.

\subsection{10x Xenium Mouse Brain Dataset}\label{10xX}

\begin{figure}[h!]
\begin{center}
{{\includegraphics[scale=0.42]{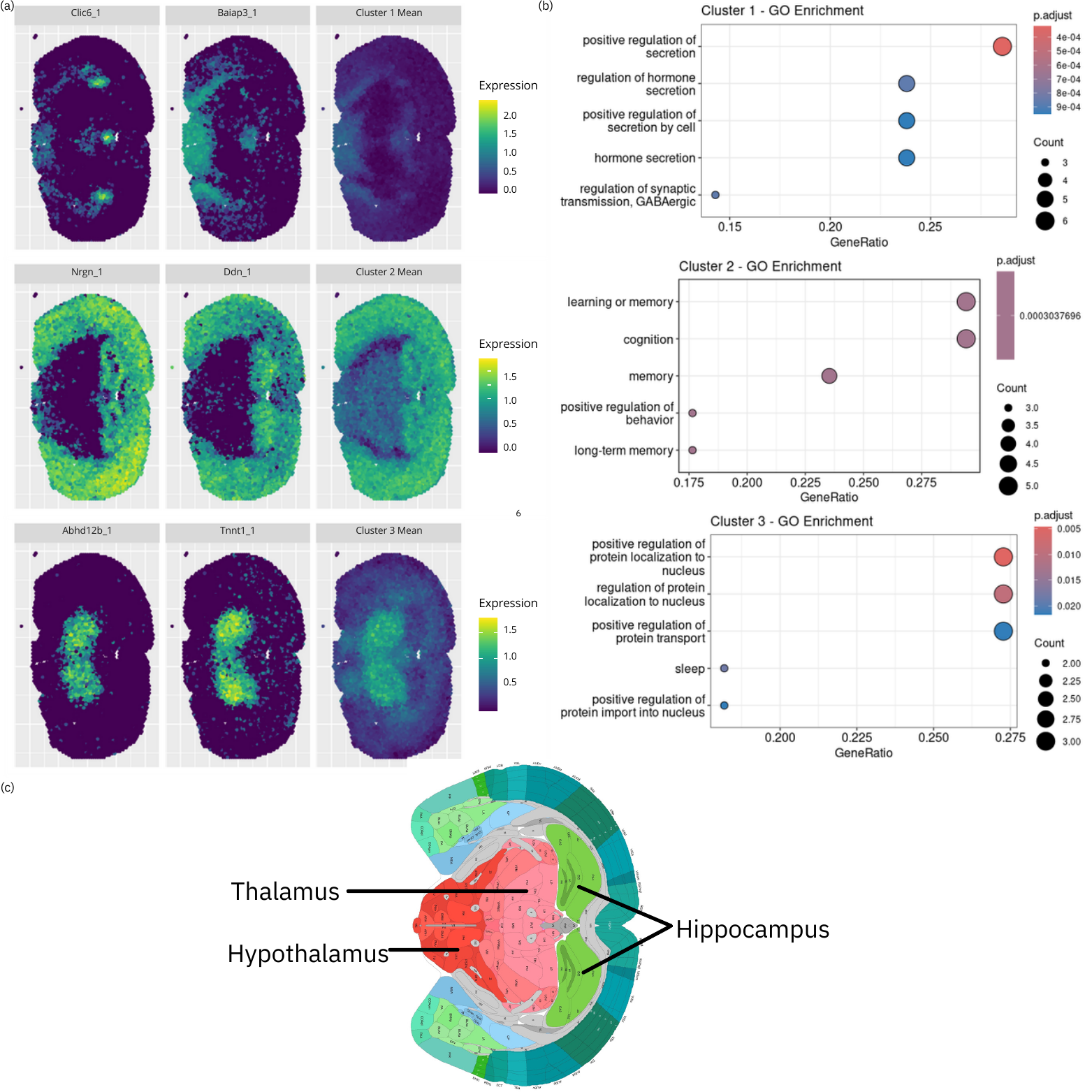}}}%
\caption{(a) Mean spatial expression patterns of the three clusters with two representative genes per cluster in the 10x Xenium mouse brain dataset. (b) GO enrichment terms for the clusters. (c) Anatomical regions from the Allen Reference Atlas – Mouse Brain (atlas.brain-map.org) \citep{allen2}. \label{Fig7}}
\end{center}
\end{figure}

The 10x Xenium mouse brain dataset was similarly processed to include 50 SVGs (Appendix \ref{secA1}). stIHC identified three co-expression modules containing 21, 17, and 12 genes, respectively. Figure \ref{Fig7}(a) shows the mean spatial expression patterns of the clusters alongside two representative genes per cluster. The clear spatial distinctions reflect the effectiveness of stIHC in capturing spatial structures, even with different sequencing technologies and tissue orientations.

GO enrichment analysis revealed distinct biological functions for each cluster (Figure \ref{Fig7}(b)), aligning with anatomical regions from the Allen Mouse Brain Atlas (Figure \ref{Fig7}(c)).
\begin{itemize}
    \item Cluster 1 (21 genes): Hypothalamus, enriched for hormone regulation processes \citep{chrousos1995hypothalamic}.
    \item Cluster 2 (17 genes): Hippocampus, associated with learning and cognition \citep{pang2019ammon,borzello2023assessments}.
    \item Cluster 3 (12 genes): Thalamus, linked to sleep regulation \citep{jan2009role}.
\end{itemize}
Despite differences in spatial resolution, molecular profiling, and tissue preparation between the 10x Visium and 10x Xenium platforms, stIHC consistently identified key regions (hippocampus, thalamus, hypothalamus) with highly similar biological functions from both datasets. GO enrichment results for the hippocampus and hypothalamus clusters aligned across both platforms, reflecting learning and memory processes in the hippocampus and hormone regulation in the hypothalamus. These findings underscore the robustness and applicability of stIHC across ST platforms, demonstrating its ability to uncover biological phenomena beyond platform-specific artifacts.

\subsection{10x Visium Human Lung Cancer Dataset}\label{10xvisBC}

\begin{figure}[h!]
\begin{center}
{{\includegraphics[scale=0.45]{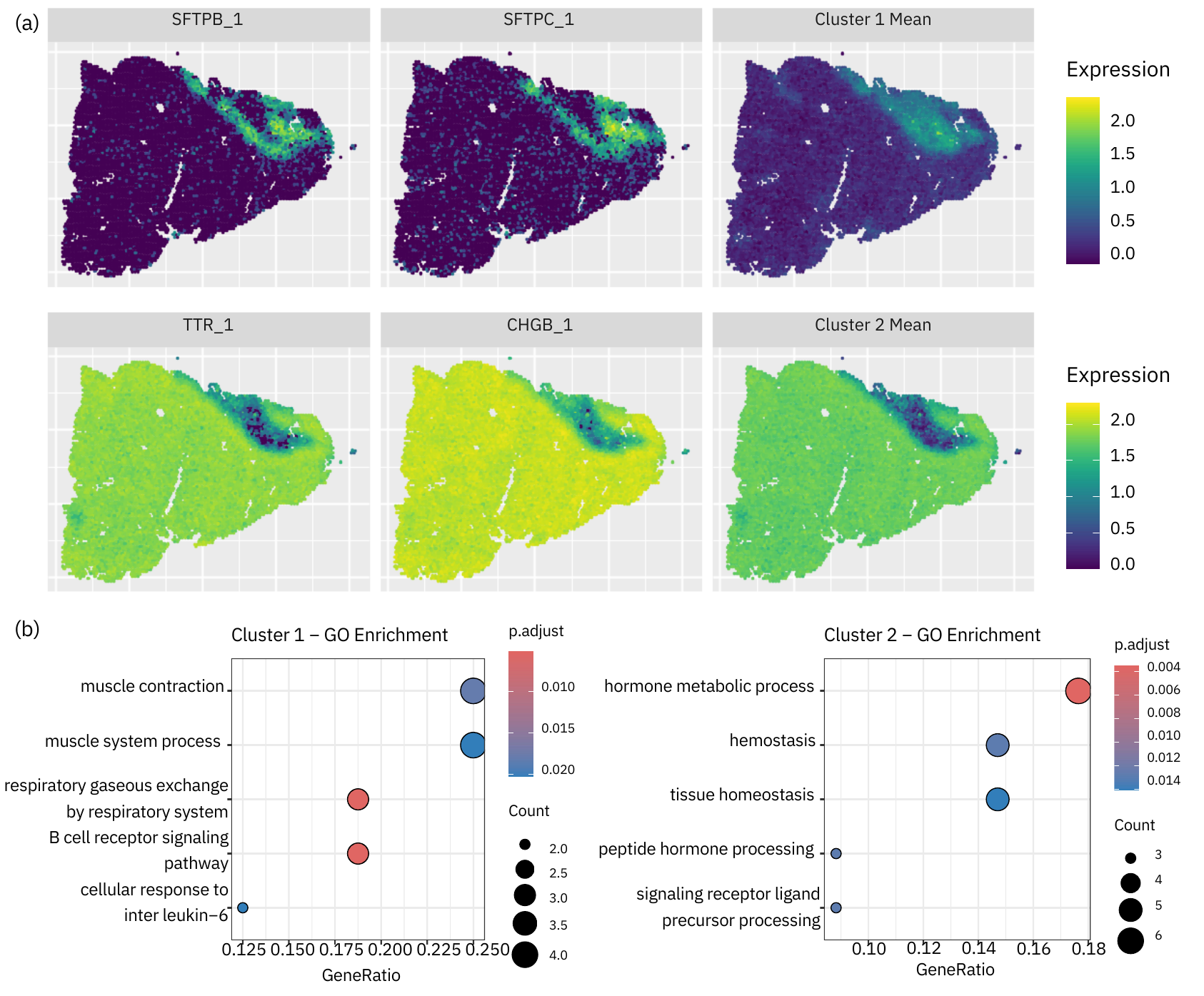}}}%
\caption{(a) Mean spatial expression patterns of the two clusters with two representative genes per cluster in the 10x Visium lung cancer dataset. (b) GO enrichment terms for the clusters. \label{figlung}}
\end{center}
\end{figure}

\noindent The 10x Visium human lung cancer dataset underwent similar processing to include 50 SVGs (Appendix \ref{secA1}). Using stIHC, two co-expression modules were identified, consisting of 16 and 34 genes, respectively. Figure \ref{figlung}(a) presents the mean spatial expression pattern for each cluster, accompanied by two representative genes from each cluster. These distinct spatial patterns illustrate that stIHC successfully captures the underlying structure of the data. We performed GO enrichment analysis and identified unique biological functions associated with each cluster (Figure \ref{figlung}(b)).
\begin{itemize}
    \item Cluster 1 (16 genes): Associated with primary lung functions, as indicated by enrichment in muscle contraction, muscle system processes, and respiratory gaseous exchange by the respiratory system. Additionally, it is linked to the immune response, as evidenced by enrichment in the B cell receptor signaling pathway, which plays a role in the tumor microenvironment \citep{leong2021b}. Furthermore, it includes genes involved in the cellular response to interleukin-6 (IL-6), which has been associated with lung cancer progression, resistance to antitumor therapies, and poor survival in lung cancer patients \citep{qu2015interleukin}. 
    
    \item Cluster 2 (34 genes): Associated with pathways that support tumor growth and development, including peptide hormone processing, which plays a known role in lung cancer development \citep{yamaguchi1986peptide}; hemostasis, as dysregulation of the hemostatic system is common in lung cancer and serves as a prognostic indicator \citep{unsal2004prognostic}; and tissue homeostasis, which contributes to tumor growth and progression \citep{quail2013microenvironmental}.
\end{itemize}

\noindent The region in the top right corner, where Cluster 1 is highly expressed and Cluster 2 is lowly expressed, likely corresponds to a functionally distinct lung region enriched in primary lung functions and immune-related processes. This aligns with the biological role of Cluster 1 genes in muscle contraction, respiratory function, and immune processes, suggesting that this area may represent non-tumor tissue or a microenvironment supportive of normal lung function, in contrast to tumor-associated regions where Cluster 2 is more highly expressed. These results indicate that the spatially distinct gene modules identified by stIHC are associated with unique biological functions. Furthermore, stIHC demonstrates the capability to uncover spatial gene co-expression modules with potential disease-specific relevance.

\newpage
\section{Application of stIHC to Mouse Olfactory Bulb Data}\label{sec5}

\begin{figure}[h!]
\begin{center}
    \includegraphics[scale=0.018]{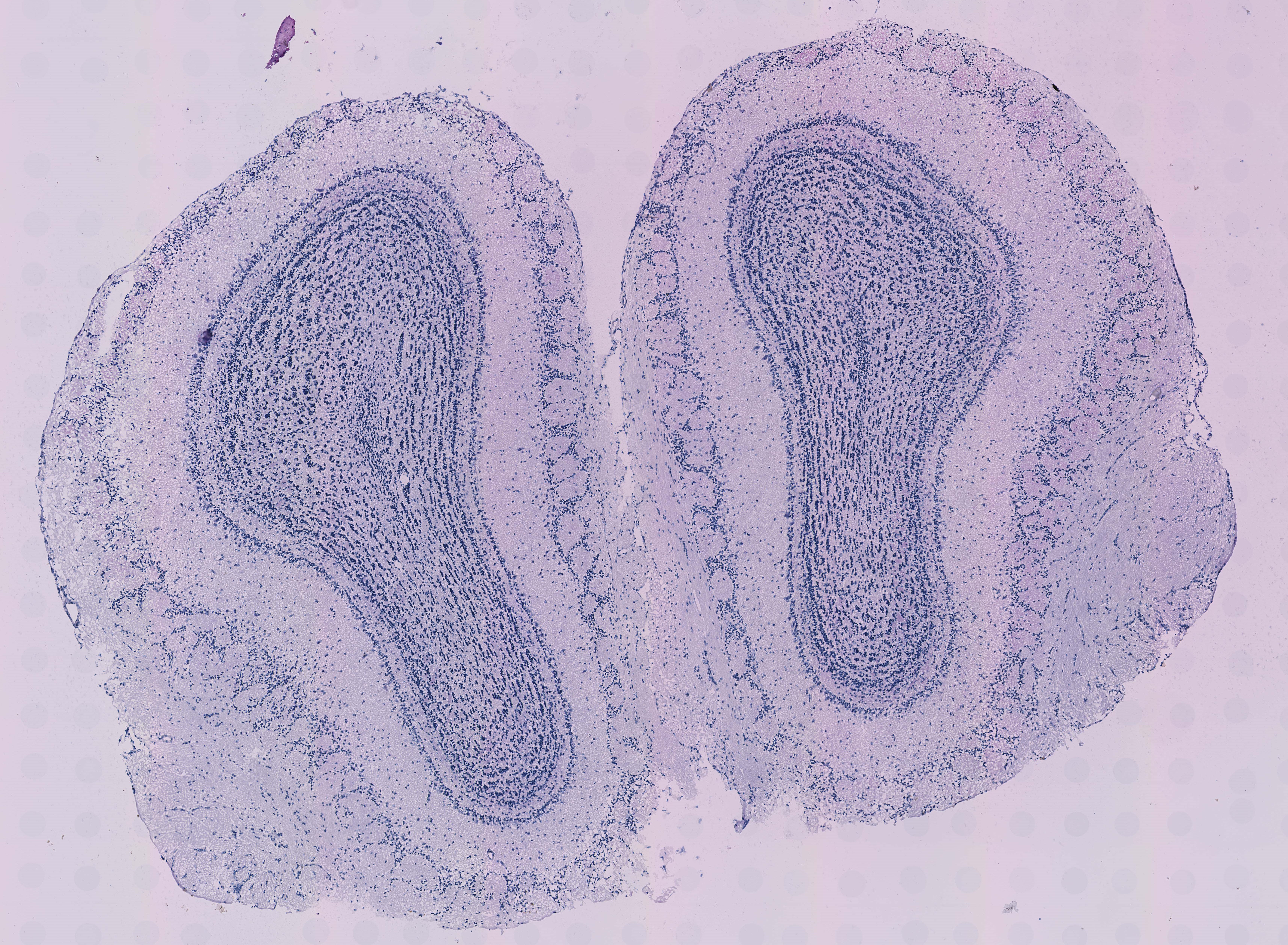}
    \caption{Hematoxylin and eosin-stained brightfield image of the mouse olfactory bulb. Image from \url{www.spatialresearch.org}.}
    \label{he}
\end{center}
\end{figure}

To demonstrate the practical application of stIHC, we analyzed ST data from replicate 11 of the mouse olfactory bulb, as described in \cite{staahl2016visualization}. This dataset, widely studied in previous research \citep{svensson2018spatialde, sun2020statistical, miller2021characterizing}, is publicly available at \url{www.spatialresearch.org}. Figure \ref{he} shows a hematoxylin and eosin-stained brightfield image of the mouse olfactory bulb, providing a structural reference for the tissue. In this analysis, we applied two preprocessing pipelines and SVG detection methods—MERINGUE and SpatialDE—to assess the flexibility and robustness of stIHC across different analytical workflows.

\subsection{Data Preprocessing and Detection of SVGs}

The raw dataset comprised 262 probe spots and 15,928 genes. Data preprocessing and SVG identification were conducted using the guidelines of MERINGUE and SpatialDE:
\begin{itemize}
\item MERINGUE: Probe spots with fewer than 100 detected genes were excluded, retaining 260 spots. Genes with fewer than 100 total reads were removed, resulting in 7,365 genes. Raw counts were normalized to counts per million (CPM) values.
\item SpatialDE: Genes with fewer than three total counts were filtered out. Variance stabilization was performed using Anscombe’s transformation to meet SpatialDE’s assumption of normally distributed data. Library size differences were adjusted by regressing out total counts for each gene. After preprocessing, 14,859 genes and 260 spots remained.
\end{itemize}
MERINGUE detected 886 SVGs, whereas SpatialDE identified 67 SVGs, highlighting substantial differences in SVG detection rates. While this study focuses on these two methods, stIHC is compatible with any SVG detection method and adaptive to diverse pipelines.

\subsection{Clustering SVGs into Co-Expression Modules}

The log-transformed counts (\( \log(1 + x) \)) of the detected SVGs were input into stIHC to cluster the genes into co-expression modules based on their spatial expression patterns. Unlike methods such as SpatialDE, SPARK, and SPARK-X, which require users to predefine the number of clusters, stIHC determines the optimal number of clusters automatically using an internal metric.

For the 886 SVGs detected by MERINGUE, stIHC identified three clusters containing 523, 341, and 22 genes. For the 67 SVGs detected by SpatialDE, stIHC identified two clusters containing 38 and 29 genes.

\begin{figure}[h!]
\begin{center}
    \includegraphics[scale=0.45]{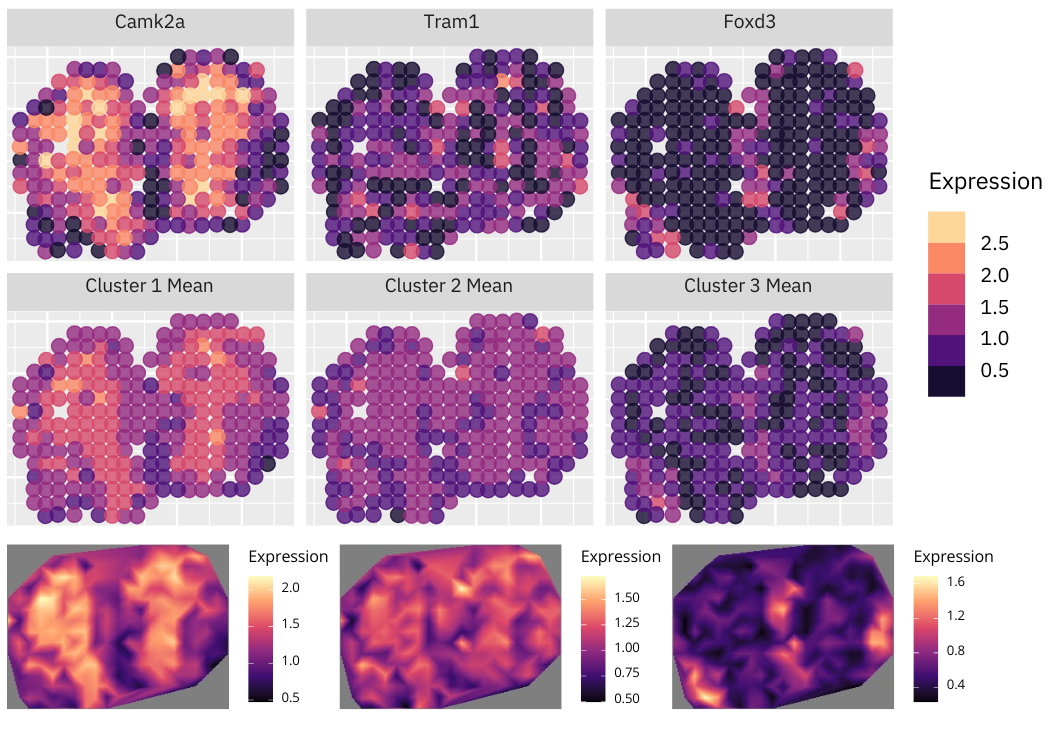}
    \caption{Three co-expression modules identified by stIHC from 886 SVGs detected by MERINGUE. Top row: representative genes for each cluster. Middle row: mean spatial expression patterns in the original resolution. Bottom row: smoothed spatial surfaces.}
    \label{dots}
\end{center}
\end{figure}

Figure \ref{dots} illustrates the spatial expression patterns of the three clusters identified by stIHC from the 886 SVGs detected using MERINGUE. The top row displays a representative gene from each cluster, while the middle and bottom rows illustrate the mean spatial expression patterns in the original spatial resolution and as a smoothed surface, respectively. 
\begin{itemize}
    \item Cluster 1: Genes in this cluster exhibited high expression localized in the granular cell layer, consistent with prior findings \citep{sun2020statistical}.
    \item Cluster 2: This cluster included genes with elevated expression in the central region but lacked a distinct spatial pattern.
    \item Cluster 3: Genes in this cluster showed high expression at the edges of the bulb, corresponding to the glomerular layer \citep{sun2020statistical}.
\end{itemize}
The spatial patterns of Clusters 1 and 3 align with findings reported in \cite{svensson2018spatialde, sun2020statistical, miller2021characterizing}.

\begin{figure}[h!]
\begin{center}
    \includegraphics[scale=0.6]{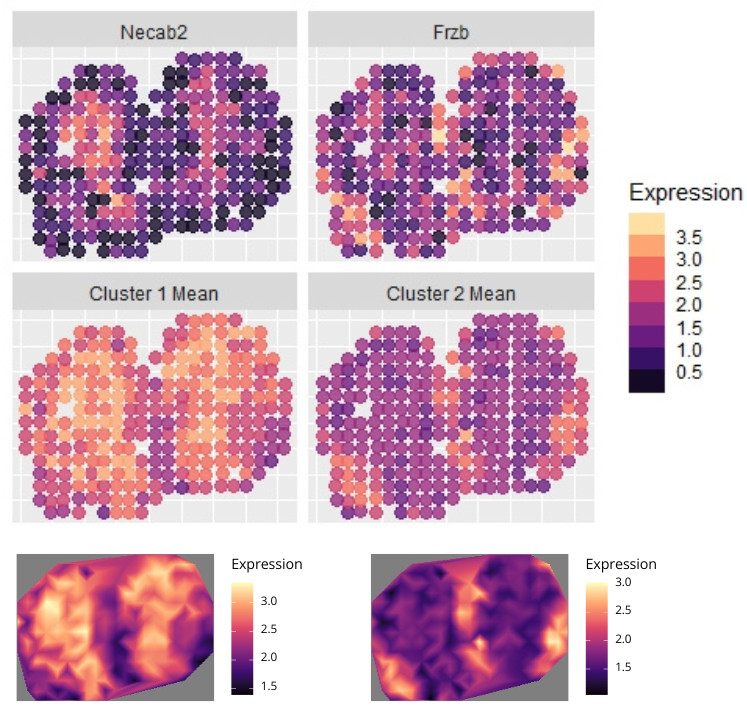}
    \caption{Two co-expression modules identified by stIHC from 67 SVGs detected by SpatialDE. Top row: representative genes for each cluster. Middle row: mean spatial expression patterns in the original resolution. Bottom row: smoothed spatial surfaces.}
    \label{dots2}
\end{center}
\end{figure}

Figure \ref{dots2} illustrates the two co-expression modules identified by stIHC from the 67 SVGs detected by SpatialDE:
\begin{itemize}
    \item Cluster 1: Genes in this cluster displayed high expression in the inner region, corresponding to the granular cell layer (analogous to MERINGUE’s Cluster 1).
    \item Cluster 2: This cluster showed high expression at the bulb's edges, aligning with MERINGUE’s Cluster 3 (glomerular layer).
\end{itemize}
The absence of a cluster corresponding to MERINGUE’s Cluster 2 suggests that the larger SVG set identified by MERINGUE may include genes with less distinct or noisy spatial patterns, as Cluster 2 does not exhibit a clear spatial structure.

These results highlight the flexibility and robustness of stIHC in identifying meaningful spatial co-expression modules from SVGs detected using different methods. Despite variations in preprocessing pipelines, the number of detected SVGs, and the resulting number of co-expression modules, stIHC consistently uncovered biologically relevant spatial patterns. Notably, stIHC preserved key spatial features, such as the granular and glomerular layers, across both MERINGUE and SpatialDE analyses, demonstrating its effectiveness in capturing the functional organization of the mouse olfactory bulb.

\section{Conclusion}\label{sec6}

We presented a novel clustering method, stIHC, designed to identify spatial co-expression gene modules from ST data. stIHC demonstrated significant advantages over existing methods in simulations, especially in detecting rare and unique spatial co-expression modules often overlooked by standard clustering approaches. While traditional methods perform adequately in datasets with balanced cluster sizes, they falter when clusters are imbalanced. By accommodating imbalanced clusters, stIHC consistently outperformed other methods, successfully identifying spatial co-expression modules even when represented by only a few genes.

When applied to ST datasets from the 10x Visium, 10x Xenium mouse brain datasets, \ch{and 10x Visium Human Lung Cancer dataset}, stIHC effectively identified biologically meaningful co-expression modules. These modules exhibited distinct spatial expression patterns \ch{and, in the case of the mouse brain datasets,} corresponded to known anatomical regions of the brain, supporting the hypothesis that spatially organized gene expression reflects the tissue’s functional structure. Gene Ontology enrichment analysis confirmed that genes within the same co-expression module shared coherent and distinct biological functions, further reinforcing the biological relevance of spatially co-expressed genes.

We evaluated stIHC's robustness and generalizability across multiple datasets and ST technologies, including 10x Visium, 10x Xenium, and Spatial Transcriptomics. As demonstrated in Section \ref{sec4}, stIHC consistently identified spatially co-expressed gene modules aligned with known biological pathways and anatomical structures, regardless of the technology or tissue orientation.

As shown in Section \ref{sec5}, stIHC can be integrated with any SVG detection method, offering a data-driven and parameter-lean clustering approach that eliminates the need for user-defined input parameters. Its ability to preserve spatial patterns while detecting biologically and functionally relevant clusters makes it a powerful tool for exploring the complex spatial organization of tissues. The R implementation of stIHC is freely available at \url{www.github.com/CatherineH1/stIHC}.

\backmatter

\bmhead{Acknowledgments}

This research was supported by Science Foundation Ireland through the SFI Centre for Research Training in Genomics Data Science under Grant numbers 18/CRT/6214 and 18/CRT/6214(S4). For the purpose of Open Access, the author has applied a CC BY public copyright license to any Author Accepted Manuscript version arising from this submission.

We thank Zhijian Li and Luca Pinello of the Broad Institute of MIT and Harvard for providing the data used in Section \ref{sec4}.

\section*{Statements and Declarations}

All authors declare that they have no competing interests. 


\clearpage
\begin{appendices}
\section{Convergence of the IHC algorithm}\label{con}
The within-cluster correlation quantifies the degree of similarity among curves within the same cluster. Conversely, the between-cluster correlation evaluates the distinctiveness of the averages of curves across different clusters. The objective is to achieve a within-cluster correlation that surpasses the predefined threshold $\alpha_{u}$, while simultaneously maintaining a between-cluster correlation that remains below $\alpha_{u}$. However, enforcing these dual conditions may lead to convergence challenges, attributable to their inherently conflicting nature. To address this, Step iv iterates through Steps ii and iii until one achieves convergence in the within-cluster correlation. Subsequently, Step v relaxes the stringent criterion on the within-cluster correlation. As a result, the clusters formed typically exhibit an average within-cluster correlation that is similar to $\alpha_{u}$, diverging from the more rigid requirement of consistently exceeding $\alpha_{u}$. For a comprehensive description of the IHC algorithm, the reader is referred to \cite{carey2016correlation}.

\section{Computational efficiency}\label{time}
Table \ref{compeff} summarizes the execution time in seconds for each method across the three simulation scenarios. Simulation 3.1, which involves 100 genes across 2,696 spatial spots, is the most computationally demanding. This is followed by Simulation 3.2 with 49 genes across 2,696 spatial spots, and Simulation 3.3 with 49 genes over 260 spatial spots. SPARK/SPARK-X is significantly faster than the other methods because, although they employ a modeling approach for detecting SVGs, they apply clustering directly to the raw gene expression data, which has been log-transformed and standardized to have a mean of zero and a standard deviation of one across all spots. In contrast, stIHC, MERINGUE, and SpatialDE perform clustering on the modeled gene expression data. This step is crucial because raw gene expression data often contain substantial noise, which can negatively impact clustering accuracy. By modeling gene expression, this noise is reduced, leading to more robust gene clusters that are less affected by noise in gene expression measurements. While SPARK/SPARK-X is the fastest method, it fails to detect the smallest cluster in each of the simulations. However, longer computation time does not necessarily yield better clustering results, as demonstrated by SpatialDE, which has the longest runtime but performs the worst across the simulations. Overall, stIHC strikes a balance between efficiency and performance, achieving optimal clustering results while maintaining reasonable computational demands. The evaluation was conducted on a personal laptop with an 11th Gen Intel Core i7-1165G7 CPU (4 cores), 32GB RAM, and an Intel Iris Xe Graphics GPU. The system was equipped with a 1TB NVMe SSD (Samsung PM9A1).

\begin{longtable}{|p{3.2cm}|p{3cm}|p{3cm}|p{3cm}|}
\caption{\ch{Time in seconds to run each of the clustering methods.}} \label{compeff}\\
\hline
\textbf{\ch{Method}} & \textbf{\ch{Simulation 3.1}} & \textbf{\ch{Simulation 3.2}} & \textbf{\ch{Simulation 3.3}}  \\ \hline
\endfirsthead
\hline
\textbf{Method} & \textbf{Simulation 3.1} & \textbf{Simulation 3.2} & \textbf{Simulation 3.3} \\ \hline
\endhead
\hline
\endfoot
\hline
\endlastfoot
\ch{stIHC} &  \ch{127s} &  \ch{63s}   &  \ch{3s} \\ \hline
\ch{MERINGUE} & \ch{4005s} & \ch{1019s} & \ch{2s} \\ \hline
\ch{SpatialDE} & \ch{4278s} & \ch{3470s}  &  \ch{552s} \\ \hline
\ch{SPARK/SPARK-X} & \ch{0.34s} & \ch{0.11s} & \ch{0.03s} \\ \hline

\end{longtable}

\section{Full list of genes from identified co-expression modules}\label{fullgenelists}
The full list of genes and associated GO terms are presented for 10x Visium mouse brain, 10x Xenium mouse brain, \ch{and 10x Visium human lung cancer} analyzed in Section \ref{sec4}. Corresponding brain regions are presented for the 10x Visium mouse brain and 10x Xenium mouse brain.

\begin{longtable} {|p{1.1cm}|p{0.9cm}|p{5cm}|p{4.8cm}|p{2.1cm}|}
\caption{10x Visium mouse brain dataset cluster summary with gene information, GO Terms, and brain regions}\label{tab1} \\
\hline
\textbf{Cluster} & \textbf{\# of Genes} & \textbf{Gene Names} & \textbf{GO Terms} & \textbf{Brain Region} \\ \hline
\endfirsthead
\hline
\textbf{Cluster} & \textbf{\# of Genes} & \textbf{Gene Names} & \textbf{GO Terms} & \textbf{Brain Region} \\ \hline
\endhead
\hline
\endfoot
\hline
\endlastfoot
1 & 2 &  \textit{Gm5741}, \textit{Pou4f1}   & equilibrioception,

suckling behavior,

peripheral nervous system neuron development,

peripheral nervous system neuron differentiation,

trigeminal nerve development & Thalamus \\ \hline
2 & 10 & \textit{Camk2n1}, \textit{Nrgn}, \textit{Slc17a7}, \textit{Hpca}, \textit{Nptxr}, \textit{Olfm1}, \textit{Ddn}, \textit{Cck}, \textit{Itpka}, \textit{Camk2a} & regulation of postsynaptic membrane neurotransmitter receptor levels, 

neurotransmitter receptor localization to postsynaptic specialization membrane,

protein localization to postsynaptic specialization membrane,

protein-containing complex localization,

memory & Hippocampus \\ \hline
3 & 16 & \textit{Pmch}, \textit{Agrp}, \textit{Hcrt}, \textit{Fezf1}, \textit{Gal}, \textit{Baiap3}, \textit{Sparc}, \textit{Nnat}, \textit{Bc1}, \textit{Resp18}, \textit{Gpx3}, \textit{Arhgap36}, \textit{X6330403K07Rik}, \textit{Hap1}, \textit{Agt}, \textit{Irs4} & feeding behavior,

drinking behavior,

regulation of behavior,

regulation of hormone secretion,

neuropeptide signaling pathway & Hypothalamus \\ \hline
4 & 7 & \textit{Mbp}, \textit{Mobp}, \textit{Enpp2}, \textit{Prkcd}, \textit{Pcp4}, \textit{Tcf7l2}, \textit{Ptgds} 
 & phospholipid catabolic process,
 
membrane lipid catabolic process,

sphingolipid catabolic process,

negative regulation of cell adhesion,

regulation of oligodendrocyte differentiation
& Thalamus, Hippocampus \\ \hline
5 & 15 & \textit{Ttr}, \textit{Defb11}, \textit{Kcne2}, \textit{Slc16a8}, \textit{Slc4a5}, \textit{Wfdc2}, \textit{X2900040C04Rik}, \textit{Kcnj13}, \textit{X1500015O10Rik}, \textit{Folr1}, \textit{Prr32}, \textit{Calml4}, \textit{Clic6}, \textit{Tmem72}, \textit{Spink8} & monoatomic anion transport,

potassium ion import across
plasma membrane & Cortical subplate \\ \hline
\end{longtable}

\begin{longtable}{|p{1.1cm}|p{0.9cm}|p{5cm}|p{4.8cm}|p{2.1cm}|}
\caption{10x Xenium mouse brain dataset cluster summary with gene information, GO Terms, and brain regions} \label{tab2}\\
\hline
\textbf{Cluster} & \textbf{\# of Genes} & \textbf{Gene Names} & \textbf{GO Terms} & \textbf{Brain Region} \\ \hline
\endfirsthead
\hline
\textbf{Cluster} & \textbf{\# of Genes} & \textbf{Gene Names} & \textbf{GO Terms} & \textbf{Brain Region} \\ \hline
\endhead
\hline
\endfoot
\hline
\endlastfoot
1 & 21 & \textit{Mfrp}, \textit{Tmem72}, \textit{Ecrg4}, \textit{Dsp}, \textit{Clic6}, \textit{Baiap3}, \textit{Gal}, \textit{Gpr151}, \textit{Dlk1}, \textit{C1ql2}, \textit{Arhgap36}, \textit{Hap1}, \textit{Adora2a}, \textit{Cartpt}, \textit{AW551984}, \textit{Ndst4}, \textit{Nnat}, \textit{Lypd1}, \textit{Gabrq}, \textit{Calb2}, \textit{Gpx3}    & positive regulation of secretion,

regulation of hormone secretion,

positive regulation of secretion by cell,

hormone secretion,

regulation of synaptic
transmission, GABAergic

& Hypothalamus \\ \hline
2 & 17 & \textit{Agrp}, \textit{Camk2n1}, \textit{Slc17a7}, \textit{Nrgn}, \textit{Ddn}, \textit{Snap25}, \textit{Camk2a}, \textit{Cck}, \textit{Kcnh5}, \textit{Vxn}, \textit{Hpca}, \textit{Psd}, \textit{Olfm1}, \textit{Slc30a3}, \textit{Mef2c}, \textit{Rtn1}, \textit{Baiap2} & learning or memory,

cognition,

memory,

positive regulation of behavior,

long-term memory,

& Hippocampus \\ \hline
3 & 12 & \textit{Ttr}, \textit{Pmch}, \textit{Hcrt}, \textit{Prkcd}, \textit{Mbp}, \textit{Abhd12b}, \textit{Tnnt1}, \textit{Zic1}, \textit{Pcp4}, \textit{Rab37}, \textit{Mobp}, \textit{Tcf7l2} & regulation of protein localization to nucleus,

positive regulation of protein localization to nucleus,

positive regulation of protein transport,

sleep,

positive regulation of protein import into nucleus,

& Thalamus \\ \hline
\end{longtable}

\begin{longtable}{|p{1.1cm}|p{0.9cm}|p{6cm}|p{5.9cm}|}
\caption{\ch{10x Visium human lung cancer dataset cluster summary with gene information and GO Terms.}} \label{lunggenes}\\
\hline
\textbf{\ch{Cluster}} & \textbf{\ch{\# of Genes}} & \textbf{\ch{Gene Names}} & \textbf{\ch{GO Terms}}  \\ \hline
\endfirsthead
\hline
\textbf{\ch{Cluster}} & \textbf{\ch{\# of Genes}} & \textbf{\ch{Gene Names}} & \textbf{\ch{GO Terms}} \\ \hline
\endhead
\hline
\endfoot
\hline
\endlastfoot
\ch{1} & \ch{16} &  \ch{\textit{Sftpb}, \textit{Mgp}, \textit{Myh11}, \textit{Sftpc}, \textit{Tagln}, \textit{Igha1}, \textit{Ptgis}, \textit{Myl9}, \textit{Tpm2}, \textit{Igkc}, \textit{Acta2}, \textit{Ogn}, \textit{Thbs2}, \textit{Fgg}, \textit{Sftpa1}, \textit{Ighg1}}   & \ch{muscle contraction, 
muscle system process, 
respiratory gaseous exchange by respiratory system, B cell receptor signaling pathway, 
cellular response to inter leukin 6} \\ \hline
\ch{2} & \ch{34} & \ch{\textit{Chga}, \textit{Tff3}, \textit{Ttr}, \textit{Chgb}, \textit{Tph1}, \textit{Cartpt}, \textit{Scg5}, \textit{Cd24}, \textit{Pcsk1}, \textit{Gc}, \textit{Aldh1a1}, \textit{Pdk4}, \textit{F5}, \textit{Pebp1}, \textit{Scg2}, \textit{Pcsk1n}, \textit{Apoh}, \textit{Vstm2a}, \textit{Cxcl13}, \textit{Actg1}, \textit{Ptprn}, \textit{Hipk2}, \textit{Cga}, \textit{Serpina1}, \textit{Ckb}, \textit{Bex1}, \textit{Tm4sf4}, \textit{Cmip}, \textit{Dio2}, \textit{Aplp1}, \textit{Meis2}, \textit{Atp1b1}, \textit{Cryba2}, \textit{Syt13}} &  \ch{peptide hormone processing, 
signaling receptor ligand precursor processing, hemostasis, 
tissue homeostasis, 
hormone metabolic processes} \\ \hline

\end{longtable}

\section{Datasets}\label{secA1}

\begin{description}
\item[Section 3: 10x Visium sagital mouse brain slice dataset]
The 10x Visium sagital mouse brain slice dataset is available in the \texttt{Seurat} R package. To represent distinct spatial patterns, we selected four distinct gene expression patterns, each serving as the representative pattern for a different cluster. Using the \texttt{scDesign3} R package we simulated 25 copies of each gene, resulting in a total of 100 genes grouped into four clusters. We followed the tutorial available at \url{https://songdongyuan1994.github.io/scDesign3/docs/articles/scDesign3-spatial-vignette.html} to generate this data.
\medskip
\item[Section 4: 10x Visium, 10x Xenium Mouse Brain Data, and 10x Visium Human Lung Cancer Data] 
The data used in Section 4 was generated by Zhijian Li and Luca Pinello and is available at: \url{https://github.com/pinellolab/SVG_Benchmarking/}. scDesign3 was used to generate biologically realistic data with various spatial variability using real-world spatial transcriptomics data as a reference. The marginal distribution of expression for each gene was modeled using the function \texttt{fit\_marginal (mu\_formula =
"s(spatial1, spatial2, bs = 'gp', k = 500)", sigma\_formula = "1", family\_use = "nb")}, which fitted the data with a generalized GP model under the Negative Binomial distribution. The joint distribution of genes was modeled using the function \texttt{fit\_copula (family\_use = "nb", copula =
"gaussian")}. Next, they extracted the mean parameters for each gene across all spots, denoted by $\mu_{s}(s)$ to remove spatial correlation between the spots, generating a non-spatially variable mean function, and used the function simu\_new to generate simulation data as follows:
\[
\mu(s) = \alpha \cdot \mu_s(s) + (1 - \alpha) \cdot \mu_{ns}(s),
\]

\noindent where $\alpha$ denotes the fraction of spatial variability in simulated gene expression. For our analysis in this paper we selected the genes for which $\alpha = 1$, meaning the expressions were generated from the GP model with the same spatial variability as the reference data.
\begin{description}
   \item[Section 5: Mouse Olfactory Bulb] The mouse olfactory bulb data for the  Replicate 11 sample can be downloaded from \url{www.spatialresearch.org}
\end{description}

\section{Existing method implementation }\label{secA2}%
\begin{itemize}

    \item[] SPARK and SPARK-X: To cluster genes SPARK and SPARK-X use the hierarchical agglomerative clustering algorithm in the R package \texttt{amap} with the two optional parameters in the R function set to be Euclidean distance and Ward’s criterion, respectively. We used this function for a range of cluster values and selected the optimal clustering partition by maximizing the silhouette index.

    \medskip
    
    \item[] MERINGUE: was implemented by following the tutorial available at \url{https://jef.works/MERINGUE/mOB_analysis}

    \medskip

    \item[] SpatialDE: The Bioconductor version of SpatialDE was implemented by following the tutorial available at \url{https://www.bioconductor.org/packages/release/bioc/vignettes/spatialDE/inst/doc/spatialDE.html}

\end{itemize}
\end{description}

\end{appendices}

\clearpage
\bibliography{sn-bibliography}

\end{document}